\documentclass[twocolumn,twoside,slac_two,nofootinbib]{revtex4}
\pdfoutput=1 
\usepackage{graphicx}
\usepackage{fancyhdr}
\usepackage{graphics}
\usepackage{epstopdf}
\usepackage{textpos}
\begin{document}

%My definitions
\def\met{\ensuremath{E_{\mathrm{T}}^{\mathrm{miss}}}} 
\def\meff{\ensuremath{m_{\mathrm{eff}}}}
\def\mO{\ensuremath{m_{0}}}
\def\mOH{\ensuremath{m_{1/2}}}
\def\pt{\ensuremath{p_{\mathrm{T}}}} 
\def\et{\ensuremath{E_{\mathrm{T}}}} 
\def\eT{\ensuremath{E_{\mathrm{T}}}} 
\def\ET{\ensuremath{E_{\mathrm{T}}}} 
\def\HT{\ensuremath{H_{\mathrm{T}}}} 
\def\mt{\ensuremath{M_{\mathrm{T}}}}

\newcommand{\GeVcc}     {\ \mathrm{GeV/c^2}}
\newcommand{\GeVc}     {\ \mathrm{GeV/c}}
\newcommand{\GeV}       {\ \mathrm{GeV}}
\newcommand{\invpb}     {\ensuremath{~\mathrm{pb}^{-1}}}
\newcommand{\invfb}     {\ensuremath{~\mathrm{fb}^{-1}}}

\newcommand{\mgl}        {\ensuremath{m_{\tilde\mathrm g}}}
\newcommand{\msq}        {\ensuremath{m_{\tilde\mathrm q}}}
\newcommand{\gl}        {\ensuremath{\tilde\mathrm g}}
\newcommand{\sq}        {\ensuremath{\tilde\mathrm q}}
\newcommand{\sqb}       {\ensuremath{\bar{\tilde\mathrm q}}}
\newcommand{\qb}        {\bar{\mathrm{q}}}
\newcommand{\ch}        {\ensuremath{\tilde\chi_1^\pm}}

\def\bjet{$b$-jet}
\def\bjets{$b$-jets}
\def\ttbar{\ensuremath{t\bar t}}

%%%%%%%%%%%%%%%%%%%%%% WRITE THE TITLE HERE %%%%%%%%%%%%%%%%%%%
\title{\centering Supersymmetry Searches at the Tevatron and the LHC Collider Experiments}
%%%%%%%%%%%%%%%%%%%%%% WRITE THE AUTHOR HERE %%%%%%%%%%%%%%%%%
\author{
\centering
\includegraphics[scale=0.15]{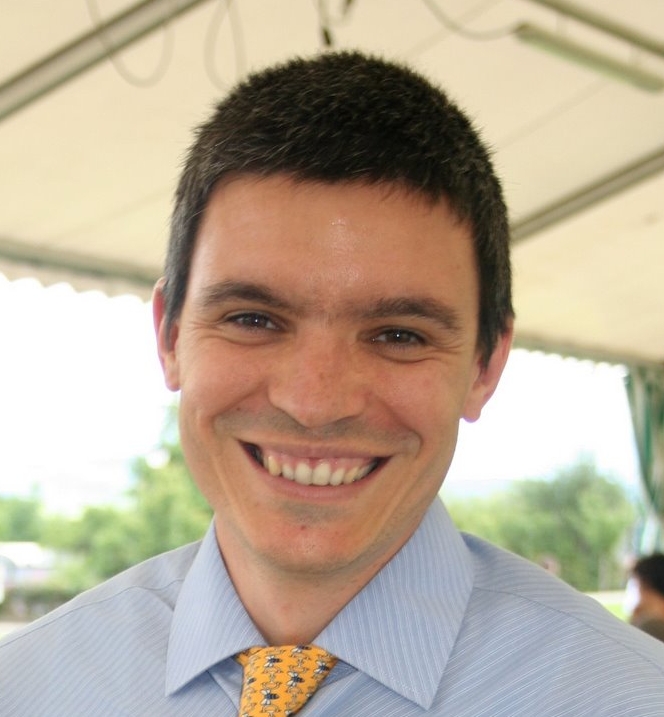} \\
\begin{center}
X. Portell Bueso{\footnote{On behalf of the ATLAS, CDF, CMS and D\O~Collaborations}}
\end{center}}
\affiliation{\centering European Organization for Nuclear Research CERN, CH-1211, Switzerland}
%%%%%%%%%%%%%%%%%%%%%% WRITE THE ABSTRACT HERE %%%%%%%%%%%%%%%%
\begin{abstract}
This document is a brief review of some of the most relevant searches for Supersymmetry carried out at the Tevatron and the LHC collider experiments, until the end of August 2011. Different final states covering $R$-parity conserving and violating scenarios have been scrutinized and no significant deviation from the Standard Model has been observed. As a result, new limits on the Supersymmetry parameter space have been established.

\end{abstract}

\maketitle
\thispagestyle{fancy}


\section{\label{sec:intro}INTRODUCTION}
Supersymmetry (SUSY) is one of the most compelling theories for physics beyond the Standard Model (SM)~\cite{susy}. It predicts a new symmetry between bosons and fermions such that for every SM particle, a superpartner should exist with a spin value differing by one half unit. This hypothesis has strong theoretical and experimental implications. On the theory side, it naturally solves the hierarchy problem~\cite{hierarchy}, a divergent value of the Higgs mass when considering radiative corrections and the SM valid up until the Planck scale. In addition, SUSY makes the unification of forces at a Grand Unification Scale (GUT)~\cite{GUT} possible. On the experimental side, the existence of several new particles, including a dark matter candidate under certain conditions, are predicted. If no particular fine tuning is introduced in the theory, these particles should be light enough to be produced at the current hadron colliders.

Since the mechanism that breaks SUSY is unknown, more than 100 new parameters are introduced in the Minimal Supersymmetric extension of the Standard Model (MSSM) to induce a soft breaking of the symmetry~\cite{mssm}. To reduce them to a more manageable set, different approaches are typically considered. The so-called ``top-down'' approach, makes some assumptions at the GUT scale and via renormalization group equations the phenomenology at the electroweak scale is predicted. CMSSM~\cite{CMSSM} or GMSB~\cite{GMSB} are among the models most commonly used in this context. Alternatively, one can follow a ``bottom-up'' approach in which different phenomenological assumptions are made at the electroweak scale to simplify the number of particles expected and their relationships. Finally, limits can also be given generically as the product of cross section, efficiency and acceptance ($\sigma\cdot\epsilon\cdot A$). In this case, it is worth mentioning that this value is provided as an upper limit on the effective cross section given the luminosity and the number of expected and observed events, without any attempt of correcting for the experimental constraints.

The Tevatron and the LHC hadron colliders are actively looking for signs of SUSY and, in their absence, constraining further the SUSY parameter space beyond the LEP legacy~\cite{LEPsusy}. Two multipurpose experiments are collecting data at each of the colliders: ATLAS and CMS at the LHC and CDF and D\O~at the Tevatron. The LHC, being a proton-proton collider currently operating at a center-of-mass energy of 7~TeV, is particularly sensitive to colored SUSY particles such as squarks and gluinos (the superpartners of the quarks and gluons, respectively), even with relatively low luminosity. The Tevatron, with a center-of-mass energy of 1.96~TeV, was the first machine establishing limits beyond LEP constraints in pair production of SUSY particles. Nowadays, it profits from the large dataset of proton-antiproton collisions to search for non-colored SUSY particles and direct production of third generation squarks, establishing the most stringent limits up to date on these processes.

The SUSY searches are generically classified in $R$-parity conserving (RPC) or violating (RPV) analyses. $R$-parity~\cite{Rparity} is a symmetry postulated to avoid some leptonic and baryonic number violating terms appearing in the SUSY superpotential. If $R$-parity is conserved, SUSY particles will always be produced in pairs and will decay in cascade until the Lightest Supersymmetric Particle (LSP) is produced. This particle is stable and constitutes a dark matter candidate, which will escape detection producing a characteristic signature of large momentum imbalance in the transverse plane (\met). On the contrary, RPV signatures are mostly characterized by the possibility of producing mass resonances from the decay of SUSY particles fully into SM particles. 

A comprehensive overview of all the different searches carried out at the Tevatron and the LHC experiments is out of the scope of this document. The reader is referred to the dedicated pages of the experiments for further information~\cite{wwwexp}. The rest of the document briefly describes the techniques and results of the different RPC and RPV searches carried out at the experiments at the Tevatron and the LHC colliders. Other more exotic scenarios, such as displaced vertices or R-hadrons, and results from indirect searches, in which experiments look for deviations of rare SM processes to constrain the SUSY parameter space, are not considered in this document.

%\subsection{\label{sec:susy}Supersymmetry}

%\subsection{\label{sec:sensitivity}Sensitivity to Supersymmetry}

\section{\label{sec:RPC}RPC ANALYSES}
The most general signature of RPC processes is the presence of large \met. In addition, SUSY cascade decays from the initial particles can be long or short and can include different number and flavor of leptons\footnote{Throughout this document, hadronically-decaying taus are considered as jets unless otherwise stated} and jets. This rich phenomenology is used by the experiments to define dedicated searches and control different type of backgrounds.

\subsection{\label{sec:nolep}Searches without Leptons}
The strong production of SUSY particles typically involves a relatively large number of jets and \met. This is one of the most characteristic signatures in SUSY models and this is why searches without leptons are the most sensitive to a large variety of scenarios. By vetoing leptons, the SM backgrounds are dominated by QCD multijet processes that have extremely large cross sections but a very small $\epsilon\cdot A$ when requiring large \met. This situation is very difficult to model with Monte Carlo (MC) simulations and different data-driven strategies are needed. Other important backgrounds are \ttbar, $W$+jet and $Z$+jet in which the $Z$ decays invisibly, constituting an irreducible background.

\begin{figure}[h!]
  \includegraphics[width=75mm]{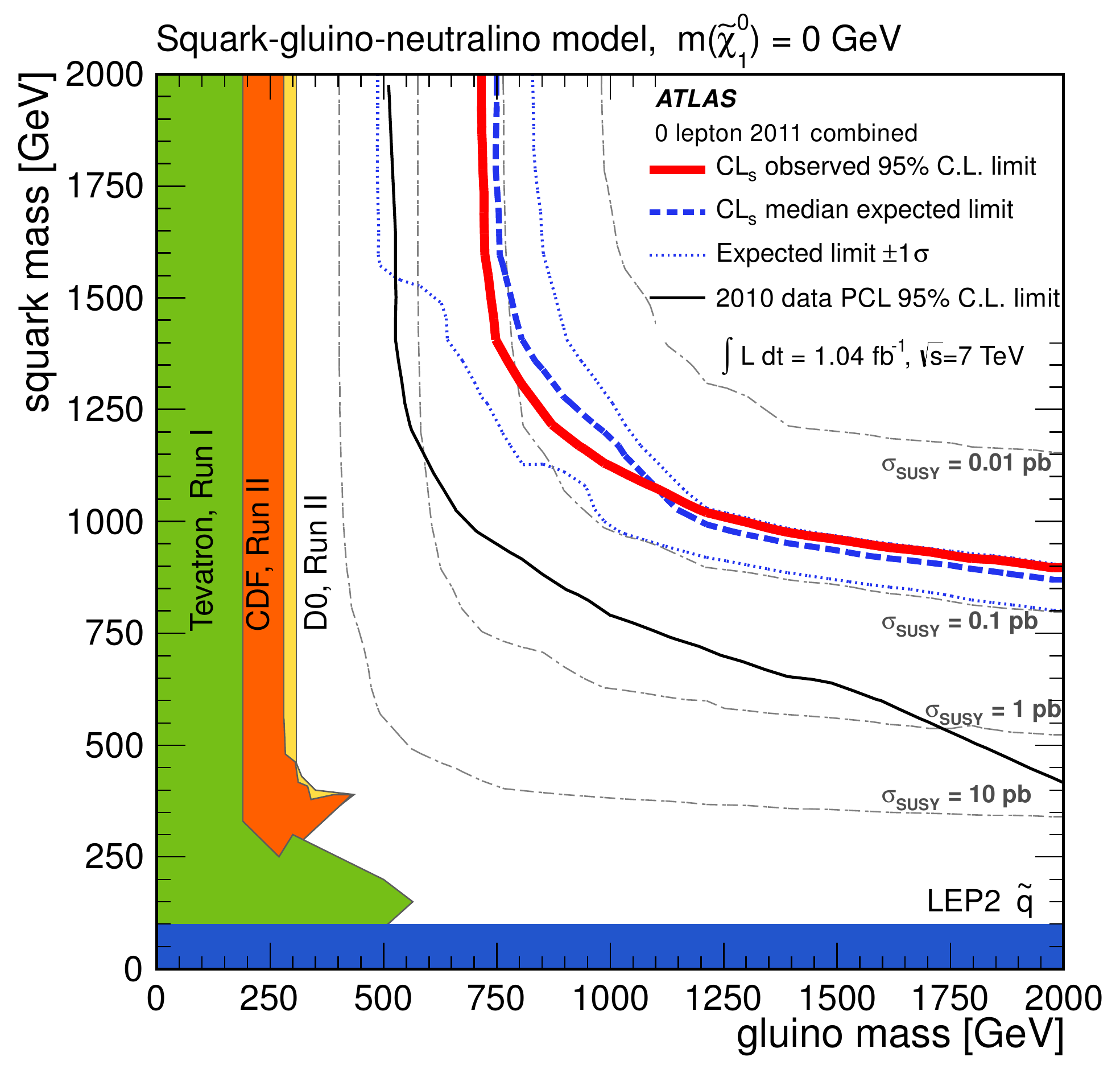}%
  \caption{\label{fig:ATLAS_0lep} ATLAS 95\% CL limits on gluino and squark masses derived from the search without leptons in a simplified model containing only gluinos, squarks from the first and second generation and a massless LSP. Previous limits are also shown for reference.}
\end{figure}

ATLAS carried out a search~\cite{ATLAS_0lep} with 1.04\invfb of integrated luminosity using the \meff~quantity, defined as the scalar sum of the \pt~of the jets and the \met. In order to maximize the sensitivity of the analysis to a variety of models, five different signal regions are defined requiring different jet inclusive multiplicities (from $\geq 2$ to $\geq 4$, with a leading jet of $\pt>130$~GeV and subleading jets of $\pt>40$~GeV), $\met>130$~GeV and different \meff~thresholds ranging from 500 to 1100~GeV. Event selections reduce the QCD multijet contribution by requiring large \met~relative to the hadronic activity and that no jet is aligned in the azimuthal plane with the \met. For each signal region, five control regions enhancing different backgrounds are defined. The QCD multijet background is estimated using a completely data-driven technique, which consists in the generation of pseudo-events following a smearing of the jets according to their response function, as derived in a low \met~significance region. This estimation is normalized appropriately in a region where at least one of the jets is aligned with the \met. The rest of the backgrounds are estimated using MC or data-driven approaches in the control regions and then a MC-driven transfer function is used to estimate the contribution in the signal region. A global likelihood fit combines all this information and takes into account the correlation between the uncertainties. No significant deviations from SM expectations are found and some limits are derived. Figure~\ref{fig:ATLAS_0lep} show the 95\% CL limits in a model with simplified phenomenology, in which all SUSY particles except for squarks of the first and second generation and gluinos are set to the 5~TeV range. In this way, the SUSY colored particles produced are forced to decay directly to the LSP, which is considered massless. These results significantly extend previous limits and are valid up to an LSP mass of 200~GeV.

A search aiming at large jet multiplicities was also carried out by ATLAS with 1.34\invfb~\cite{ATLAS_multijets}. In this case, signal regions are defined by six, seven or eight jets with \pt~ranging 55 to 80~GeV. The main background contribution is from QCD multijet production, which is controlled by using the fact that $\met/\sqrt{\HT}$ (with \HT~being the scalar \pt~sum of the jets) is invariant under jet multiplicities. This assumption was validated in many different control regions. The rest of the backgrounds are estimated using MC and validated in dedicated control regions requiring one muon. No significant deviations are found and gluino masses below 520~GeV (680~GeV under the assumption that $\msq = 2\cdot \mgl$) are excluded at 95\% CL in a CMSSM model with $\tan\beta=10$, $A_0=0$ and $\mu>0$.

\begin{figure}[h!]
  \includegraphics[width=85mm]{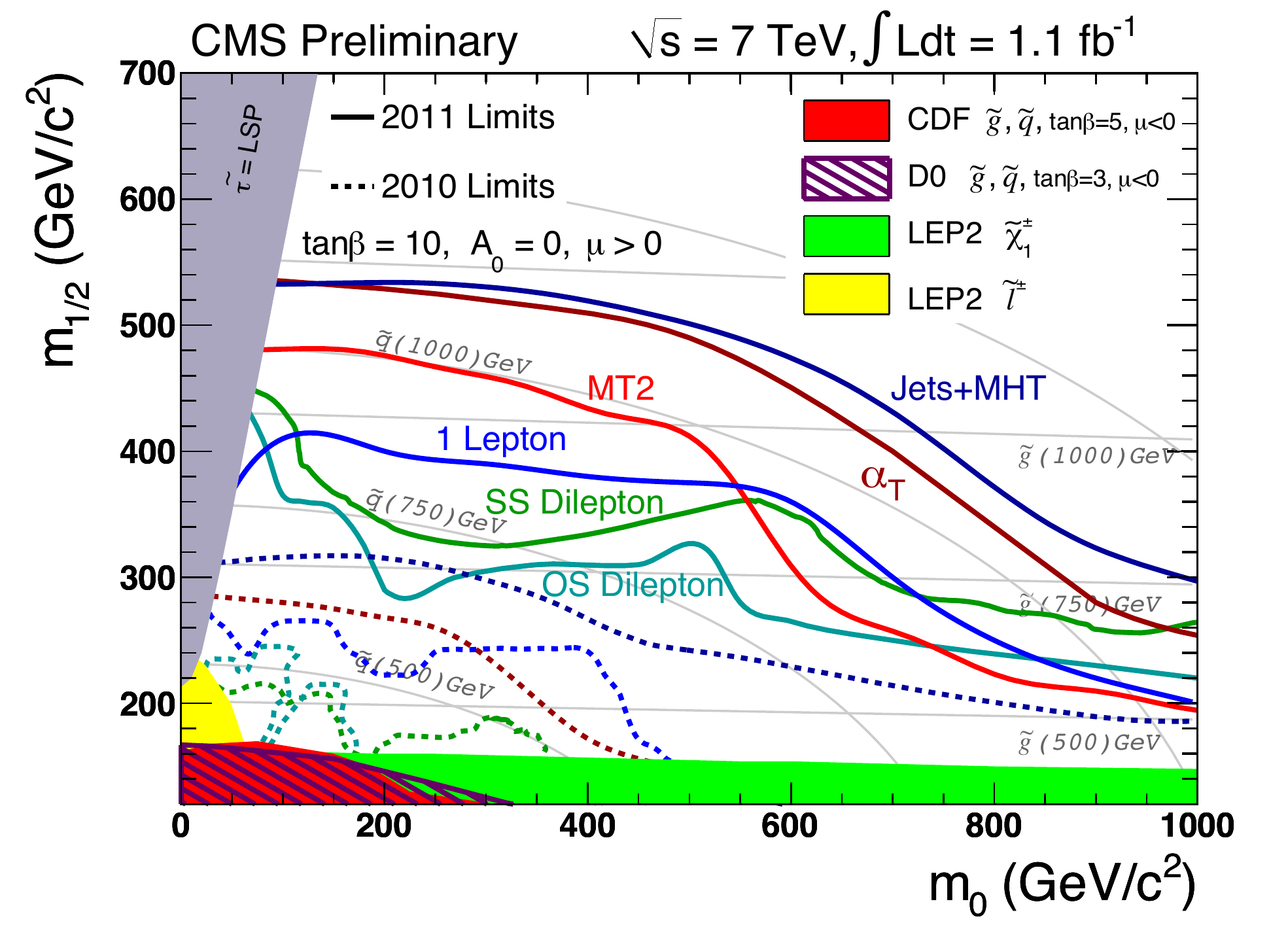}%
  \caption{\label{fig:CMS_tanb10} CMS 95\% CL exclusion limits from many different searches in a CMSSM scenario with $\tan\beta=10$, $\mu>0$ and $A_0=0$.}
\end{figure}

CMS carried out as well a series of searches aiming at the same signature but with a special focus on topological variables to discriminate against backgrounds. In the following, two of them are described: $\alpha_T$~\cite{CMS_alphaT} and Razor~\cite{CMS_Razor} searches. The $\alpha_T$ variable~\cite{alphaT} is defined as the ratio between the \pt~of the second leading jet and the transverse mass between the first two leading jets. In back-to-back topologies, such as QCD multijet production, this ratio shows a strong cutoff at 0.5, providing a good handle to discriminate against this type of background. In the case of more than two jets in the event, the two-jet topology is achieved by clustering all the jets that are relatively close in pseudo-rapidity and azimuthal distance, using a dedicated algorithm. The CMS analysis uses 1.1\invfb of data and the fact that $R_{\alpha_T}$, the ratio between events with $\alpha_T>0.55$ and $\alpha_T<0.55$, is flat versus \HT~for the SM background. This information is exploited, together with some data-driven predictions, in a global likelihood fit. The experiment uses multiple \HT~bins to maximize the sensitivity and good agreement between data and expectations is found in all of them. This result significantly extends the previous limits produced with only $35\invpb$ of data and are interpreted in a CMSSM benchmark scenario of $\tan\beta=10$, $\mu>0$ and $A_0=0$, as shown in Figure~\ref{fig:CMS_tanb10}, together with other results from different searches.

The Razor quantity~\cite{Razor} is also exploited by CMS in a dedicated analysis with 35\invpb of data. This search clusters the jets until a dijet topology is obtained and then the system is boosted back to the center-of-mass frame. The $M_R$ quantity is defined to be the momentum of the jets in this system, where both jets are equal in momentum since the pair produced SUSY particles are of the same mass. This variable is defined only from energy and $z$-momentum components and has the property to peak at the mass difference between the produced particles and the invisible particles that escape detection, with a width that relates to the initial boost from radiation. In this way, the traditional search looking for an excess at the tails of some kinematic distributions can be converted into a bump-hunting search. The transverse version of this quantity, $M_{RT}$, is also defined and enters into the razor variable definition, $R=M_R/M_{RT}$. In this way, $R$ is dimensionless and combines longitudinal and transverse information. The analysis performs a fit to evaluate the different backgrounds using some {\it ansatz} defined at dedicated control regions. The signal region is defined as $R>0.5$ and $M_R>500$~GeV and $5.5\pm 1.4$ events are expected, which is in agreement with the 7 events observed.

%==
\subsection{\label{sec:onelep}Searches with One Lepton}
\begin{figure}[h!]
  \includegraphics[width=75mm]{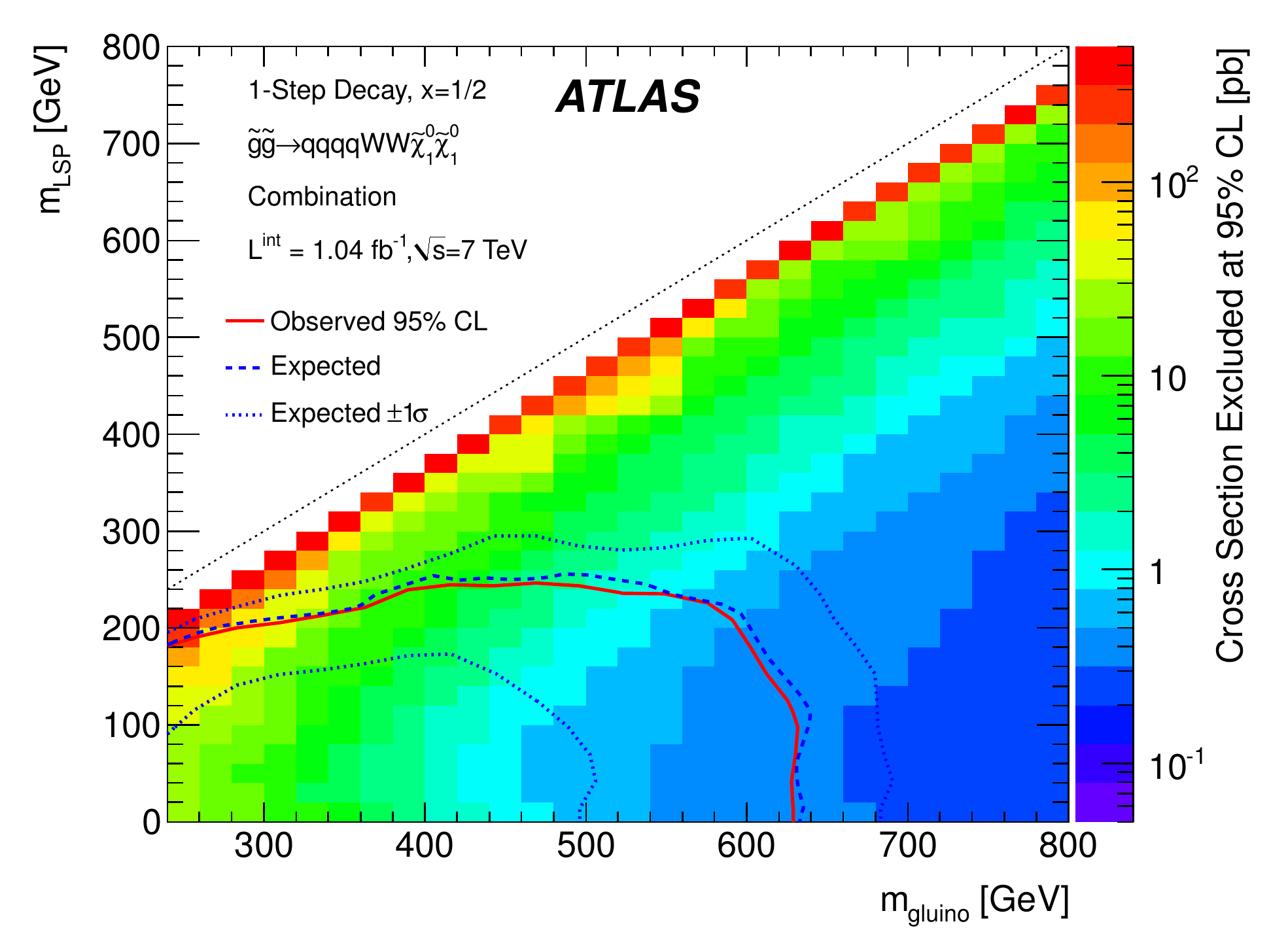}%
  \caption{\label{fig:ATLAS_1lep} ATLAS excluded cross sections at 95\% CL with a dedicated one-lepton analysis for processes in which gluinos are pair-produced and each of them decays into a quark and chargino, subsequently producing a real or virtual $W$ and the LSP. The chargino is imposed to have a mass exactly at $x = (m_{\ch} - m_{\mathrm{LSP}})/(\mgl - m_{\mathrm{LSP}})=1/2$. The solid and dashed lines are the exclusion limits when the MSSM scenario is considered.}
\end{figure}

Requiring the presence of at least one lepton in the event reduces the yield of some type of background processes, like QCD multijet production, and makes the analysis sensitive to SUSY cascade decays involving leptons. ATLAS developed a search with 1.04\invfb~\cite{ATLAS_1lep} of data in which four signal regions are defined with three or four jets in the final state and with different kinematic thresholds in order to increase the sensitivity to a generic set of models. The transverse mass between the lepton and the \met~together with the \meff~quantity, now with the lepton included in the definition, are exploited to increase sensitivity. The QCD multijet contribution is assessed in a completely data-driven manner using a matrix method~\cite{ATLAS_MM}. The rest of SM backgrounds are predicted using MC normalized to data in dedicated control regions and multiplied by a MC-driven transfer factor to estimate the corresponding contribution in the signal region. The different results and their uncertainties are finally combined in an overall likelihood fit and found to be compatible with the observed number of events. These null results are interpreted in different models, such as the one shown in Figure~\ref{fig:ATLAS_1lep}, where 95\% CL limits are derived in a simplified topology where only the gluino, the LSP and an intermediate chargino are relevant. The colored scale indicates cross sections excluded for any beyond SM process with similar topology and the lines indicate the expected and observed exclusions in the MSSM case. CMS has also recently released a one-lepton analysis~\cite{CMS_1lep} with 1.1\invfb of data in which no deviation from SM expectations is found and these results are interpreted in the context of CMSSM, as shown in Figure~\ref{fig:CMS_tanb10}. 

%==
\subsection{\label{sec:twolep}Searches with Two Leptons}
Searches with two identified leptons in the final state are also sensitive to strong production processes. Different cases depending on whether the leptons have opposite sign (OS), same sign (SS), different flavor (DF), same flavor (SF) or combinations like OSSF can be addressed and lead to different background estimation techniques.

CDF developed a SS dilepton analysis~\cite{CDF_SS}, using 6.1\invfb of data, aiming at squark or gluino pair-production with an intermediate neutralino and chargino decaying via a real or virtual $W$ or $Z$ boson. Backgrounds yields are dominated by processes containing real leptons (dibosons) and lepton misidentification from jets ($W$+jet and \ttbar) or conversions ($Z/\gamma^*$ and \ttbar). No deviations from the SM expectations are found.

CMS also developed a SS dilepton analysis with a null result using 0.98\invfb~\cite{CMS_SS} of integrated luminosity. In this case, different flavor combinations (including taus) are considered together with several \pt, \HT~and \met~thresholds. For each of the cases, a dedicated data-driven technique is used to estimate the different background contributions. The results are interpreted in terms of limits in the CMSSM scenario, as shown in Figure~\ref{fig:CMS_tanb10}. 

\begin{figure}[h!]
  \includegraphics[width=70mm]{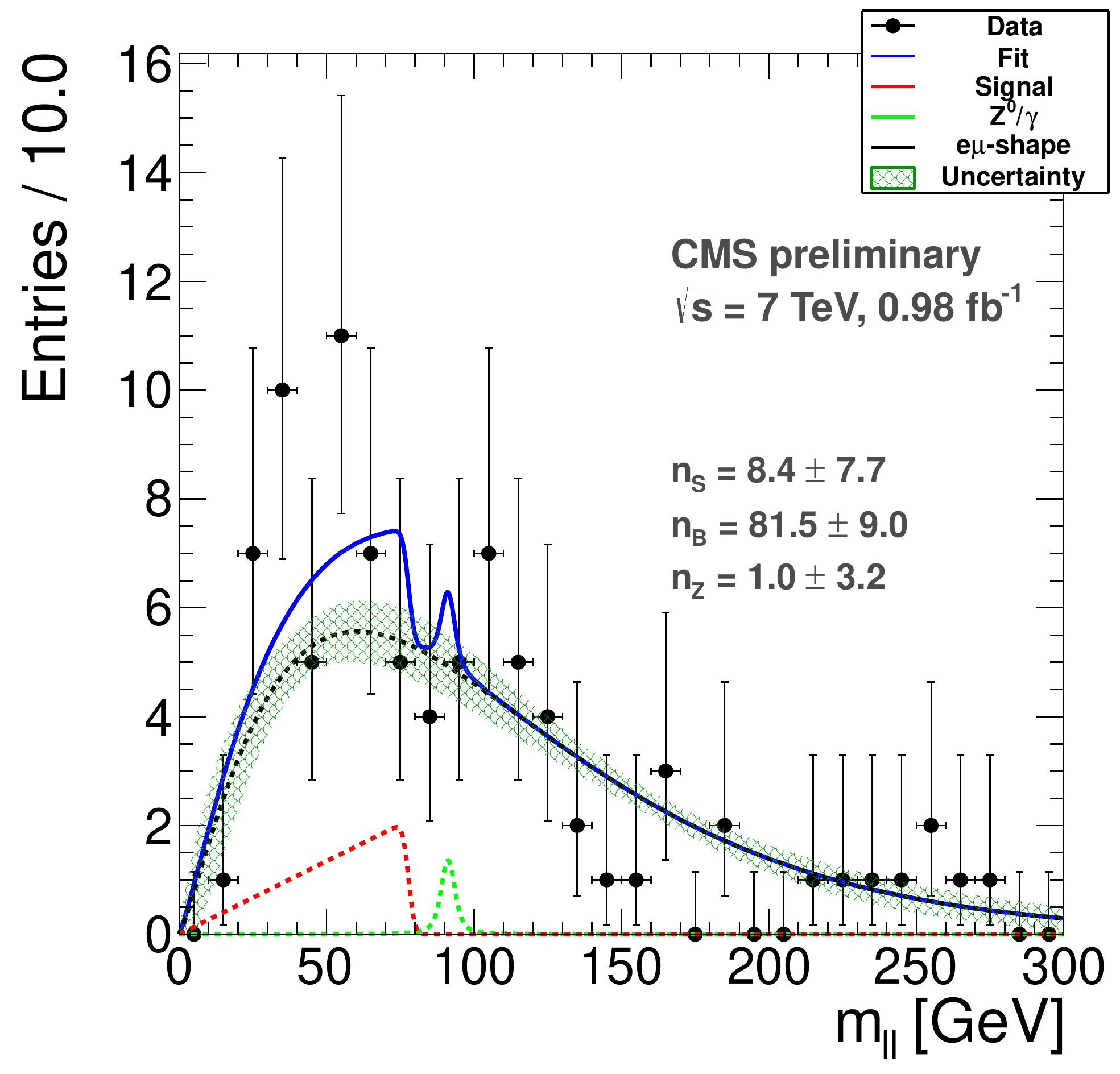}%
  \caption{\label{fig:CMS_OSSF}Results of the maximum likelihood fit to the dilepton mass distribution for events in the CMS OSSF signal region.}
\end{figure}

CMS has also released results with 0.98\invfb of integrated luminosity in a dilepton OS channel using two different approaches~\cite{CMS_OS}. The first one investigates the presence of an excess in the OSSF combination. In SUSY, cascades such as $\tilde\chi_2^0\to l\tilde l\to ll\tilde\chi_1^0$ are expected and the invariant mass of the OSSF leptons produced in this way would form a characteristic kinematic edge that relates to the mass difference between the SUSY particles. Thus, unbinned maximum likelihood fits are performed in control and signal regions, defined respectively as $100<\HT<300$~GeV and $\HT>300$~GeV. As shown in Figure~\ref{fig:CMS_OSSF}, good agreement with the expectation is observed. The other approach follows a canonical counting experiment with two different signal regions defined at high \met~or \HT~and with three different data-driven methods to estimate the backgrounds. Good agreement between observed and expected yields in all cases is found and limits are also derived in the context of CMSSM, as shown in Figure~\ref{fig:CMS_tanb10}.

ATLAS has also recently released results for OS, SS and OSSF dilepton combinations with 1\invfb\cite{ATLAS_2l} of data. For OS (SS) analyses, three (two) signal regions are defined, with at least one of them requiring large \met~and no jet requirement. For the OSSF, an excess of SF over DF is tested over a background-only hypothesis calculated with pseudo-experiments and taking into account the different uncertainties. In all cases, no excess is observed with respect to the SM expectations.

%==
\subsection{\label{sec:multilep}Searches with Multiple Leptons}
Analyses requiring three leptons in the final state are particularly sensitive to production of uncolored particles such as a chargino and neutralino, which may decay via virtual $W$ or $Z$ bosons or via sleptons, if it is kinematically allowed. SM backgrounds producing three leptons in the final state and significant \met~are small and mostly reduce to diboson production and \ttbar~with a lepton from a semi-leptonic decay of a \bjet. This final state has been considered as the golden signature for SUSY searches at the Tevatron due to the particularly favorable signal-to-background ratio. Thus, despite the fact that with a data sample $\sim 1$\invfb~the LHC may become as powerful as the Tevatron soon, the current most sensitive searches for these processes have been performed at CDF and D\O.

D\O~developed a search with 2.3\invfb of integrated luminosity in four different channels by combining electrons and muons with an isolated track and taus~\cite{D0_trilep}. The trigger performance establishes the minimum possible \pt~threshold of the objects: $\pt>(12, 8)$~GeV for two-lepton triggers and 15~GeV for single muon trigger, needed for the tau case. Two different \pt~selections per channel are implemented. An extensive set of cuts exploiting kinematic information such as invariant masses, \HT, angular distributions, etc. are applied in each of the different channels, aiming at reducing the dominant backgrounds. No significant deviation from the background expectation is observed in any of the selections.

\begin{figure}[h!]
  \includegraphics[width=80mm]{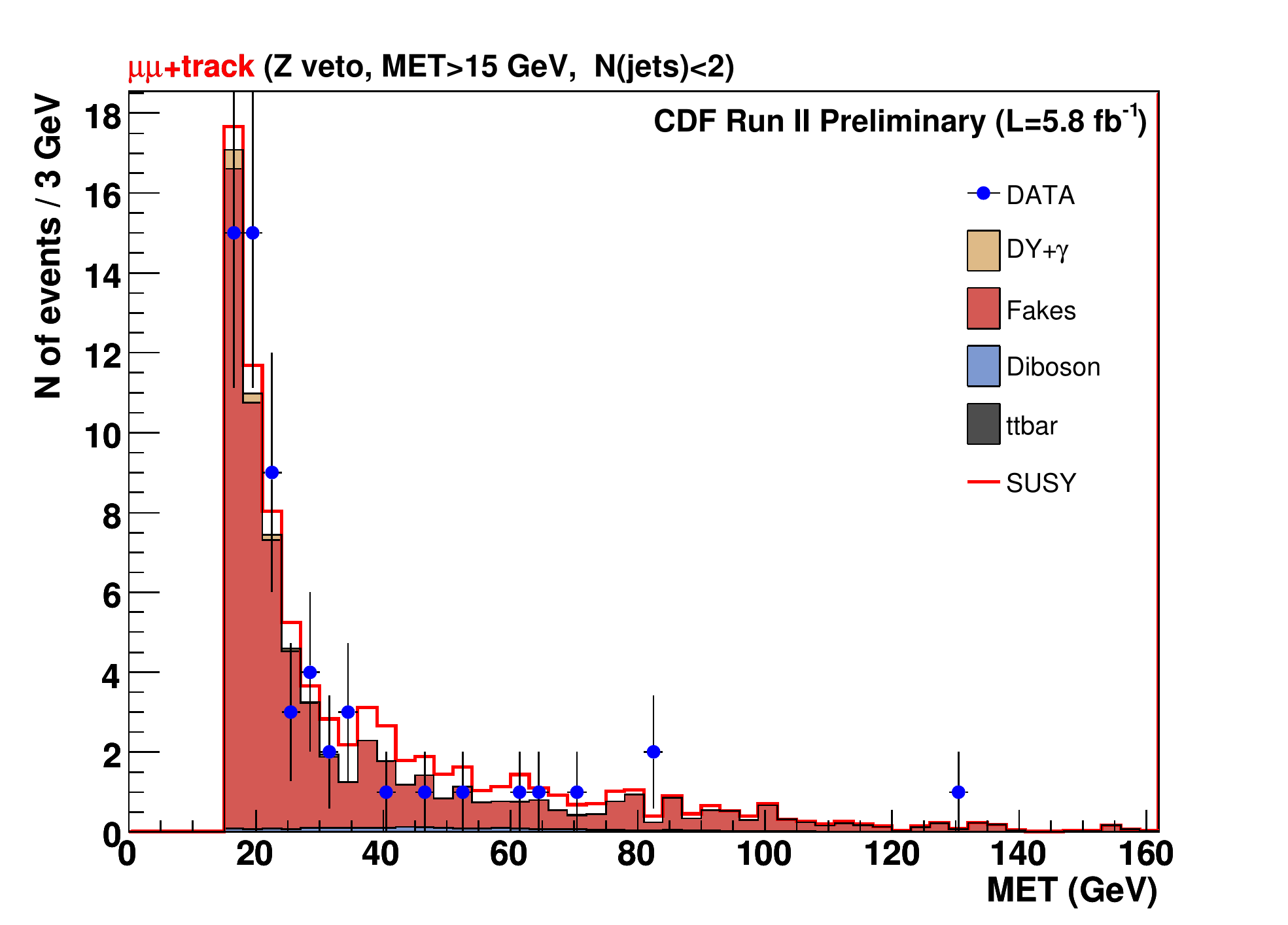}
  \caption{\label{fig:CDF_trilep}Distribution of \met~in one of the signal regions of the CDF analysis with two muons and a track.}
\end{figure}

CDF updated recently~\cite{CDF_trilep} their previous study on trileptons by considering 5.8\invfb of data and eight different exclusive channels, combining two electrons or two muons with a third object that could be an electron, a muon, a tau or a track in \pt~ranges between 5 and 20~GeV. In order to control the description of the different backgrounds, mostly dominated by Drell-Yan with a misidentified jet, 24 (40) control regions were defined in the dilepton and track (trilepton) case. As shown in Figure~\ref{fig:CDF_trilep} for the case of dimuon and track selection, no significant deviation from SM expectations is observed. CDF excludes at 95\% CL chargino mass below 168~GeV in a CMSSM scenario with \mO=60~GeV, $\tan\beta=3$, $A_0=0$ and $\mu>0$. This limit is similar to the one obtained by D\O.

%==
\subsection{\label{sec:bjets}Searches with $b$-jet Tagging}
SUSY particles of the third generation, such as the stop, the sbottom or the stau, could have significantly lower masses than the rest of the SUSY particles due to the mixing between the weak left- and right-handed eigenstates.

Searches for the direct production of sbottoms at CDF, using 2.65\invfb~\cite{CDF_sbottom} of data, and D\O, using 5.2\invfb~\cite{D0_sbottom} of data, focus on the simplified case of $\tilde b \to b+\tilde\chi_1^0$. The final state signature of two \bjets~and \met~is exploited by requiring one or two $b$-tagged jets, a lepton veto and some dedicated kinematic variables to reduce the top and QCD multijet backgrounds. One loose and one tight selections are imposed in both experiments in order to enhance the sensitivity to different $\tilde b - \tilde\chi_1^0$ mass differences. Since no deviations from expectations are observed, sbottom masses between approximately 230 and 250~GeV are excluded when the LSP mass is below 70~GeV. 

\begin{figure}[h!]
  \includegraphics[width=72mm]{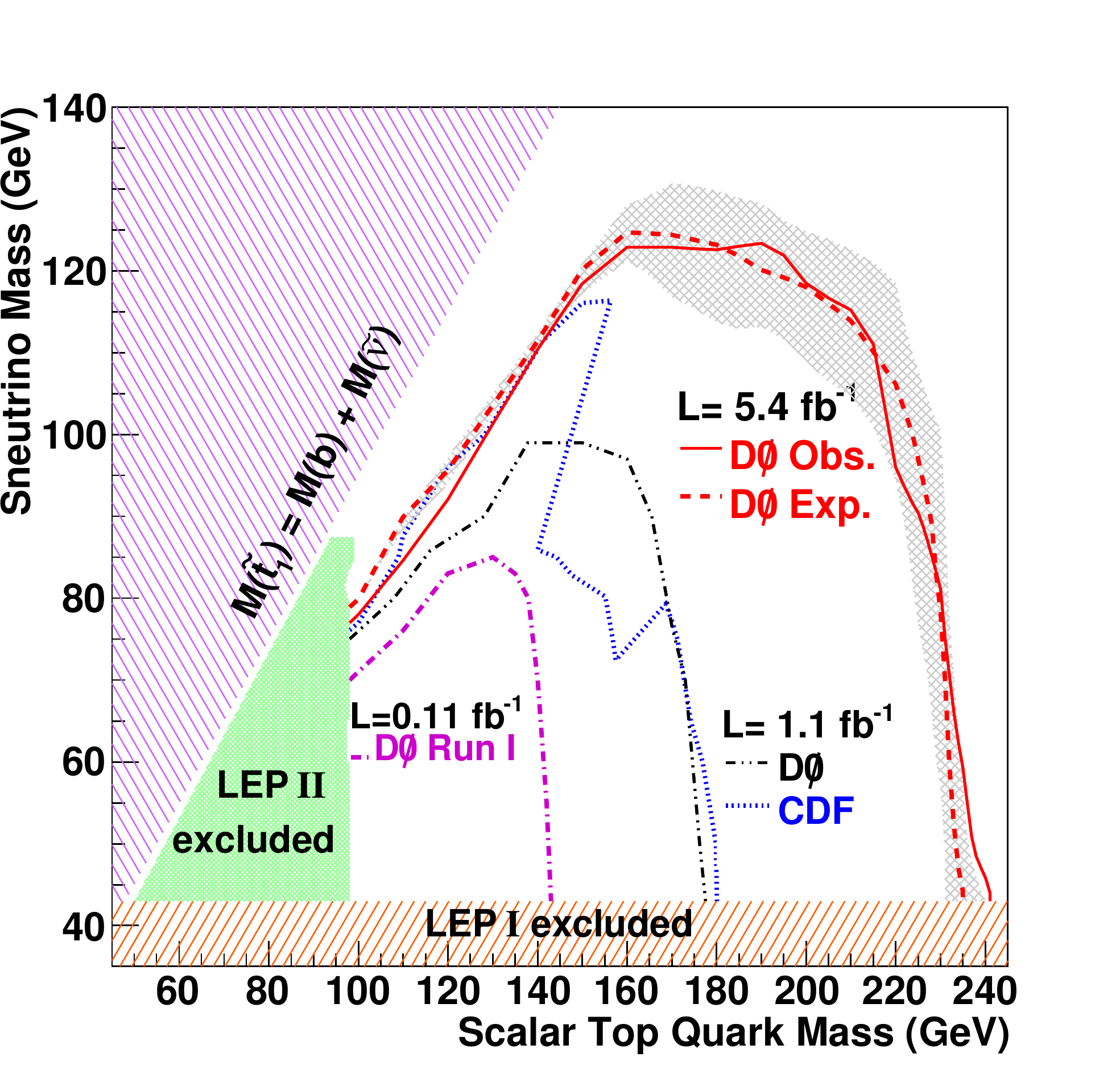}%
  \caption{\label{fig:D0_stop} Observed and expected 95\% CL exclusion regions on the scalar top mass for different sneutrino mass values in the direct stop search performed by D\O. The shaded band around the expected limit shows the effects of the scalar top quark pair production cross section uncertainty. Other limits from previous analyses are also shown for reference. }
\end{figure}

D\O~recently published a search for direct stop production with 5.4\invfb~\cite{D0_stop} of integrated luminosity. The stop can decay in many different final states depending on its own mass and that of other SUSY particles such as charginos, neutralinos and sleptons. In this analysis, the targeted scenario is a decay via a sneutrino: $\tilde t\bar{\tilde t}\to(b e\tilde\nu)(\bar b\mu\tilde\nu)$. The main backgrounds for OSDF dileptons are $Z\to\tau\tau$, dibosons and dileptonic top. A discriminant using a linear combination of different variables is built and two selections optimized for small and large stop-sneutrino mass differences are considered. Since data is found to be in agreement with the SM, limits on the stop mass as a function of the sneutrino mass are derived, significantly extending the previous results, as shown in Figure~\ref{fig:D0_stop}.

\begin{figure}[h!]
  \includegraphics[width=75mm]{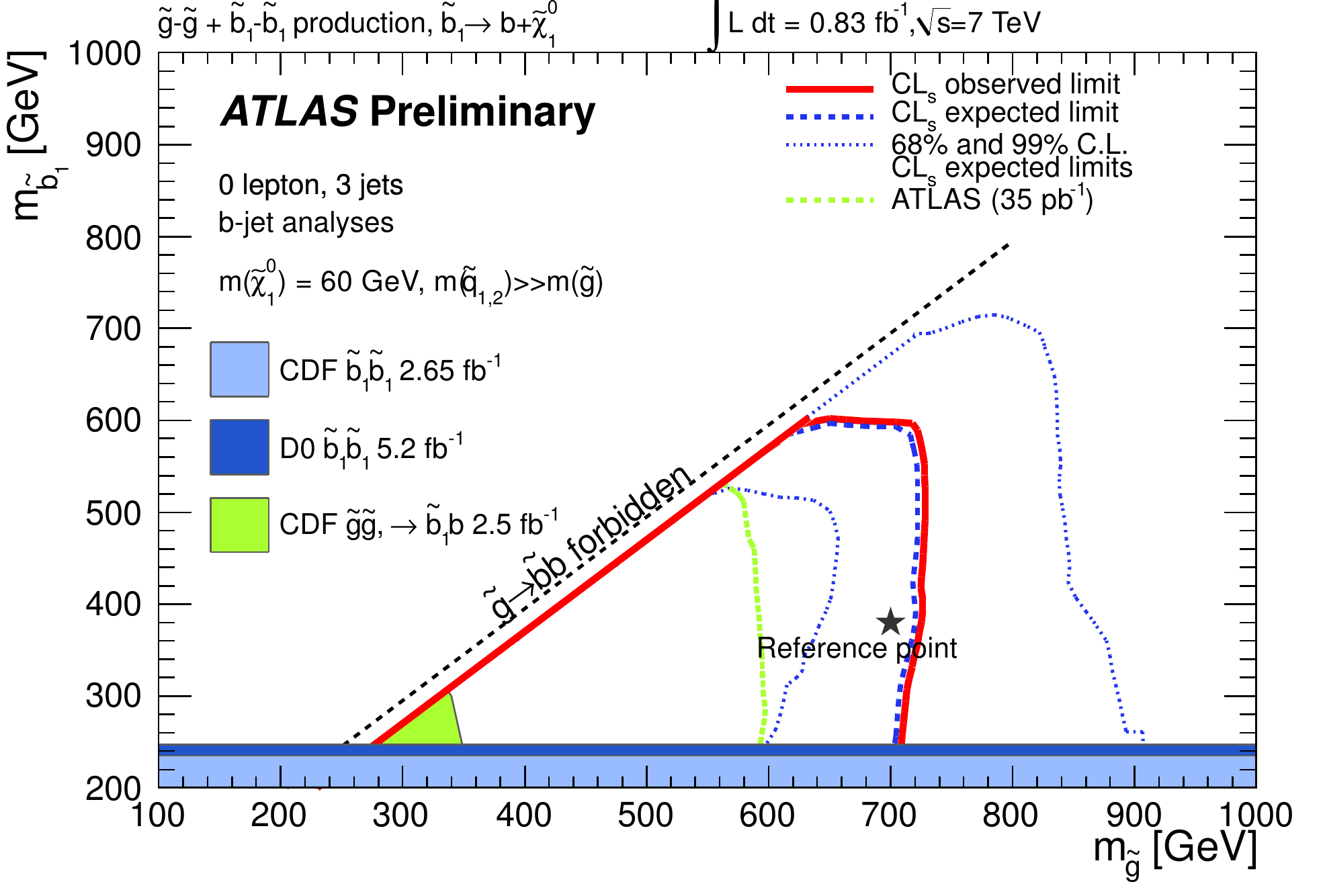}%
  \caption{\label{fig:ATLAS_glsbottom} Exclusion limits at 95\% CL in the gluino-sbottom mass plane for the ATLAS gluino-mediated sbottom production analysis. Here, the neutralino mass is set to 60~GeV and other limits are shown for reference, including the direct sbottom constraints from Tevatron in the same scenario.}
\end{figure}

Similarly to the situation in direct gaugino production searches, with a dataset of 1\invfb, the LHC is not yet as sensitive as the Tevatron in searches for direct production of third generation particles. Instead, ATLAS developed an analysis with 0.83\invfb of integrated luminosity targeting gluino-mediated production of sbottom, which has a larger cross section and provides a striking signature of four \bjets~and \met~\cite{ATLAS_glsb}. The gluino is assumed to decay via on-shell or off-shell sbottom to the LSP and all other SUSY particles are assumed to be decoupled. Four different signal regions are defined requiring either one or two $b$-tagged jets and \meff~thresholds of 500 or 700~GeV. A lepton veto is also applied and the QCD multijet background is determined fully data-driven, as in the ATLAS search without leptons described in Section~\ref{sec:nolep}. Other SM backgrounds are evaluated using MC and validated with semi data-driven estimations by requiring one lepton. No significant deviations are observed and these null results are interpreted in different theoretical models. Figure~\ref{fig:ATLAS_glsbottom} shows the extension of the limits with respect to the Tevatron and the previous ATLAS results with only 35\invpb of integrated luminosity, in the scenario in which the gluino is heavier than the sbottom and all the other SUSY particles are set at a higher scale except for the neutralino, which has a mass of 60~GeV.

In addition, ATLAS performed a gluino-mediated stop search with 1.03\invfb~\cite{ATLAS_glst} of integrated luminosity. In this case, the gluino is forced to decay to the LSP via an on-shell or off-shell stop. In the former case, the stop decays into $b\tilde\chi_1^\pm$ or $t\tilde\chi_1^0$, depending on the mass. The search is performed requiring four jets, one lepton and at least one $b$-tagged jet, as well as large \met, \meff~and transverse mass between lepton and \met. The SM expectation is estimated via fully or semi data-driven techniques to be $54.9\pm 13.6$ and 74 events are observed in data. Gluino masses are excluded approximately below 500~GeV with a small dependence on the stop mass.

%==
\subsection{\label{sec:photon}Searches with Photons}
One of the most favorable SUSY models with photons in the final state is GMSB~\cite{GMSB}. In this model, SUSY particles acquire masses via gauge interactions, which are proportional to the breaking scale $\Lambda$. In this context, the gravitino is always the LSP and different types of next-to-LSP (NLSP) can be considered. In the case of a $\tilde\chi_1^0$ NLSP being mostly bino\footnote{The SUSY partner of the U(1) gauge boson\label{fn:bino}}, the predominant decay is to a photon and a gravitino, yielding a diphoton and \met~signature. Backgrounds to this signature can be classified in QCD ``instrumental'' (mainly from diphoton, photon+jet and dijet productions), electroweak ``genuine'' ($\gamma+(W\to e\nu)$) and irreducible backgrounds ($(Z\to\nu\nu)+\gamma\gamma$ and $(W\to l\nu)+\gamma\gamma$). The two former backgrounds can be treated using data-driven techniques and the latter is usually small and assessed using MC predictions.

\begin{figure}%
  \includegraphics[width=70mm]{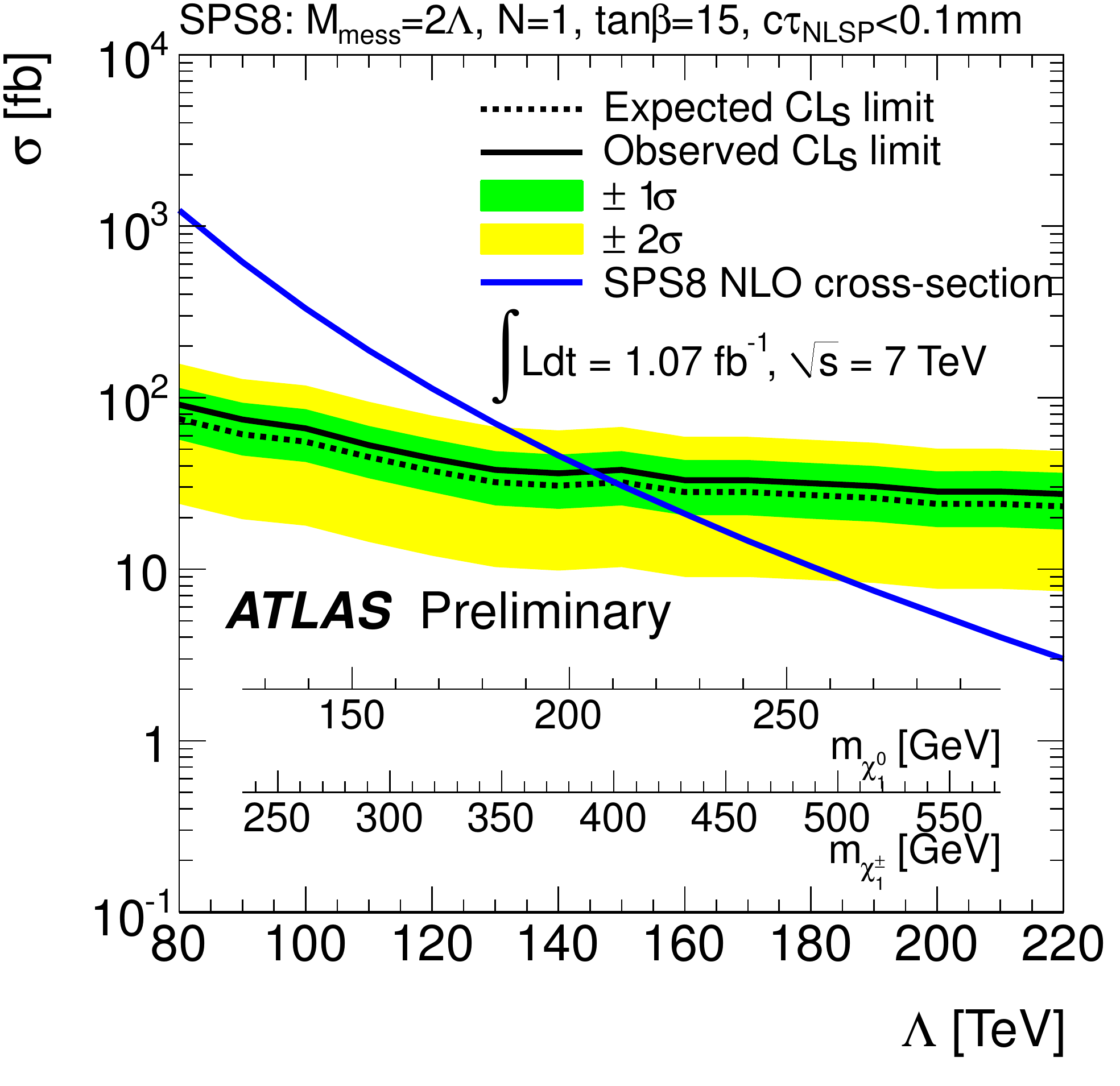}%
  \caption{\label{fig:ATLAS_SPS8}ATLAS expected and observed 95\% CL upper limits on the SPS8 production cross section as a function of $\Lambda$ and the lightest chargino and neutralino masses.}
\end{figure}

All four experiments performed a search for this final state using very similar techniques and reported null results. Tevatron searches were focused on the GMSB SPS8 scenario~\cite{SPS8}, which is dominated by gaugino pair production. D\O, with 6.3\invfb of data, excluded $\tilde\chi_1^0$ masses below 175~GeV~\cite{D0_photons} and CDF, with a smaller dataset of 2.6\invfb, constrained the NLSP masses also as a function of the NLSP lifetime~\cite{CDF_photons}. Since the LHC is more sensitive to strong production, experiments targeted the Generalized Gauge Mediated (GGM) model~\cite{GGM}, in which the constraints at the GUT scale have been relaxed to allow for almost arbitrary values of squark and gluino masses. Both ATLAS~\cite{ATLAS_photons} and CMS~\cite{CMS_photons}, with approximately 1\invfb of data, excluded squarks (gluino) masses below $\sim 700$ ($\sim 800-900$)~GeV, when assuming all other SUSY particles at higher scales. In addition, as shown in Figure~\ref{fig:ATLAS_SPS8}, ATLAS produced for the first time exclusion limits in the SPS8 scenario that extend D\O~ limits by $\sim 30$~GeV in the $\tilde\chi_1^0$ mass.

%%%%%%%%%%%%%%%%%%%%%%
\section{\label{sec:RPV}RPV ANALYSES}
$R$-parity violating terms in the SUSY lagrangian are strongly constrained by experimental limits (e.g. proton lifetime)~\cite{Rparity}. Experiments usually assume all couplings to be zero except the less constrained couplings, such as $\lambda'_{311}$ and $\lambda_{312}$, where indices refer to the family and couplings are described in the superpotential as $\lambda_{ijk}\hat L_i \hat L_j \hat E_k +\lambda'_{ijk}\hat L_i \hat Q_j \hat D_k$. Searches in RPV scenarios focus on finding a resonance produced by the decay of the SUSY particles to SM particles.

\begin{figure}[h!]
  \includegraphics[width=80mm]{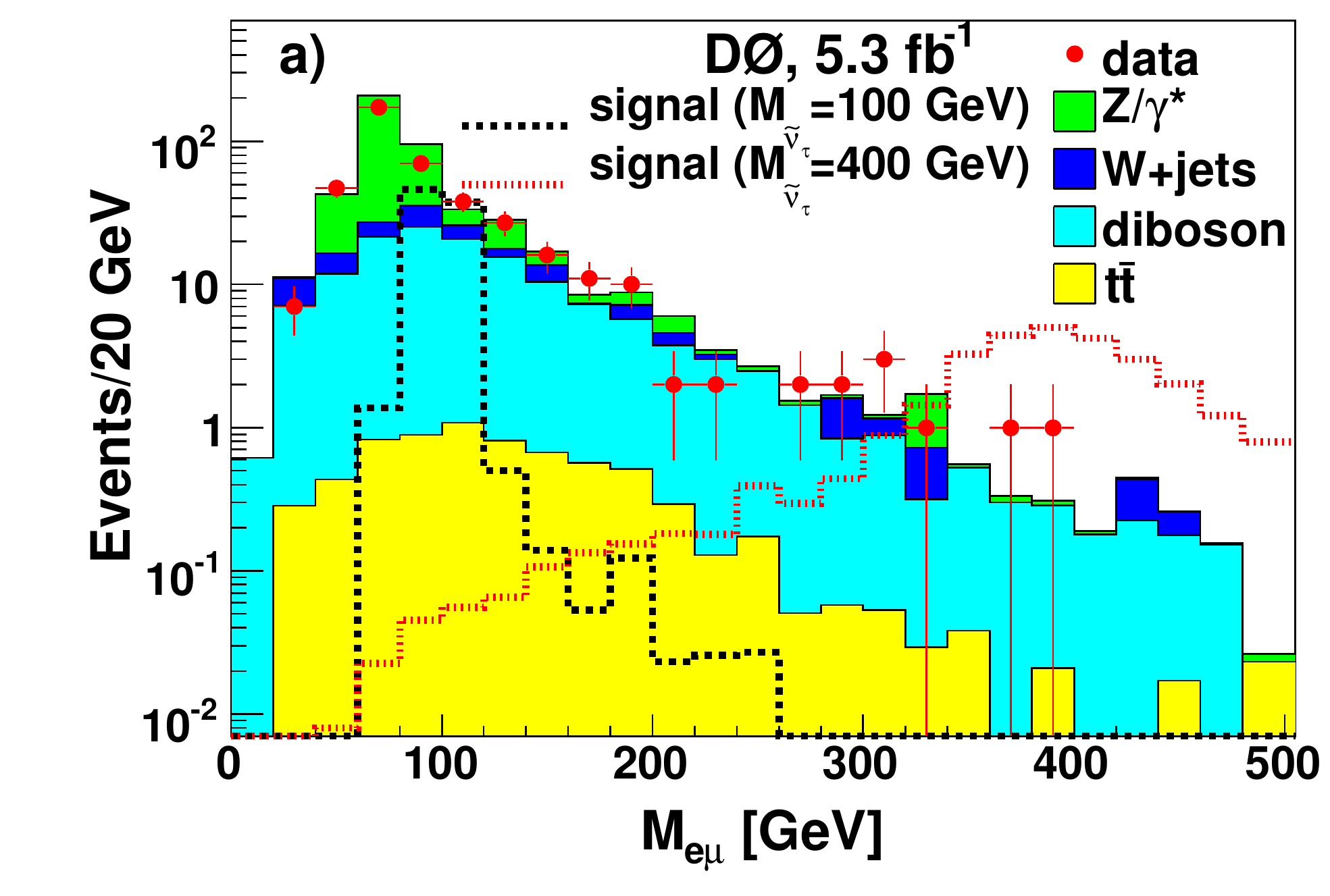}%
  \caption{\label{fig:D0_snutau}Invariant mass of $e\mu$ final states for different SM processes and two signal samples used for reference in the D\O~stau neutrino search.}
\end{figure}

%==
\subsection{\label{sec:snutau}Searches for Scalar Tau Neutrino}
\begin{figure}[h!]
  \includegraphics[width=85mm]{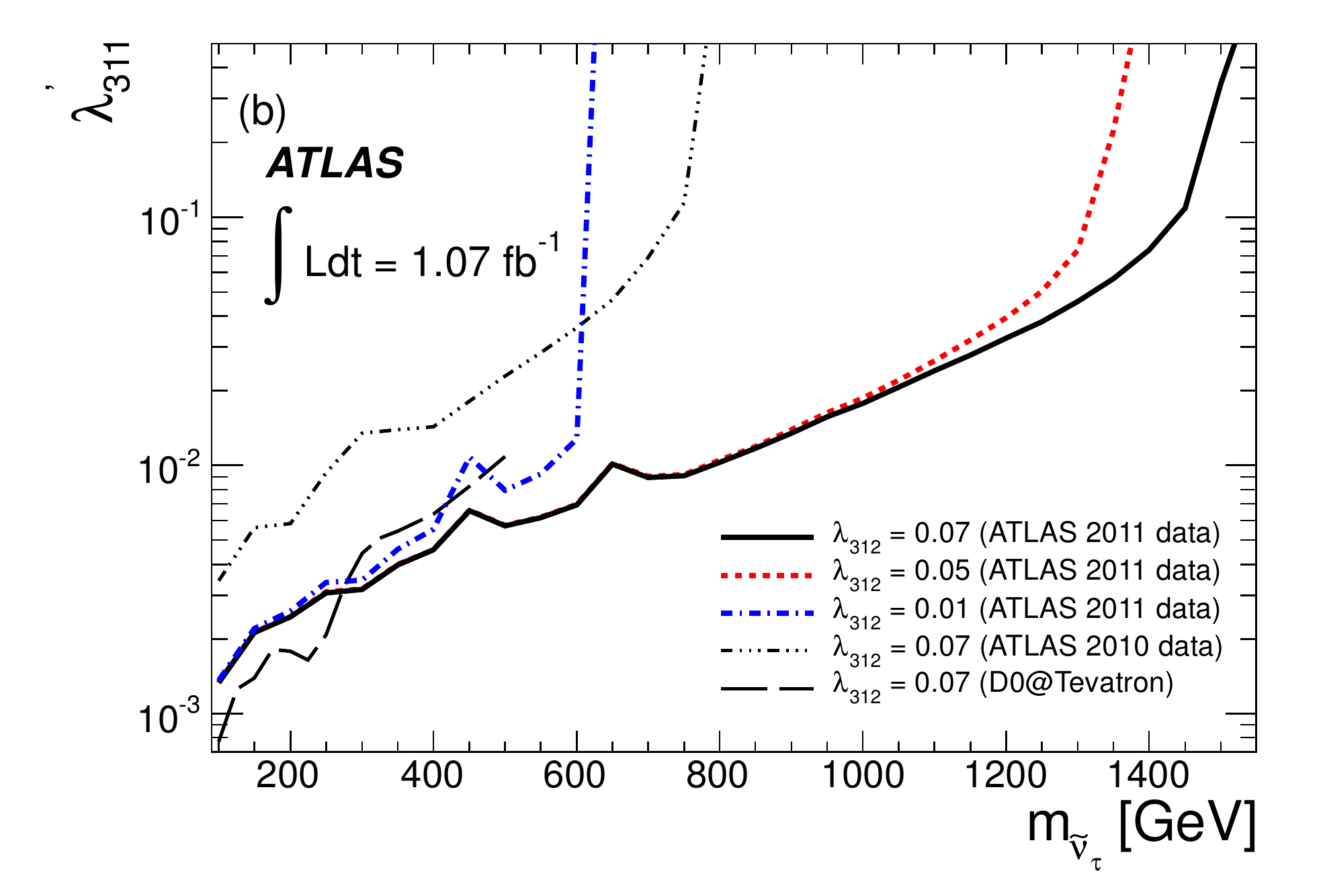}%
  \caption{\label{fig:ATLASD0_RPV}Upper 95\% CL limits on the $\lambda'_{311}$ coupling as a function of $\tilde\nu_\tau$ mass for three values of $\lambda_{312}$. Regions above the curves are excluded by either ATLAS or D\O~scalar tau neutrino searches.}
\end{figure}

A search for RPV scalar tau neutrino decaying to an electron and a muon was carried out in D\O~ using a data sample of 5.3\invfb~\cite{D0_staunu}. After some cuts to require exactly one electron and muon and to reduce the jet fake contamination, no evidence of a mass resonance peak is found, as shown in Figure~\ref{fig:D0_snutau}. A similar analysis but requiring opposite sign leptons and some different background techniques was performed by ATLAS with 0.87\invfb~of data~\cite{ATLAS_staunu}. No deviation from SM was found and limits are translated in a plane of $\tilde\nu_\tau$ production coupling ($\lambda'_{311}$) against $\tilde\nu_\tau$ mass for different decay coupling ($\lambda_{312}$) values, as shown in Figure~\ref{fig:ATLASD0_RPV}. These limits exemplify the current complementarity between the different experiments since D\O~is more competitive at lower masses whereas it is limited at higher masses, which is the region in which ATLAS is more sensitive.

%==
\subsection{\label{sec:jetresonance}Searches for Jet Resonances}
\begin{figure}[h!]
  \includegraphics[width=85mm]{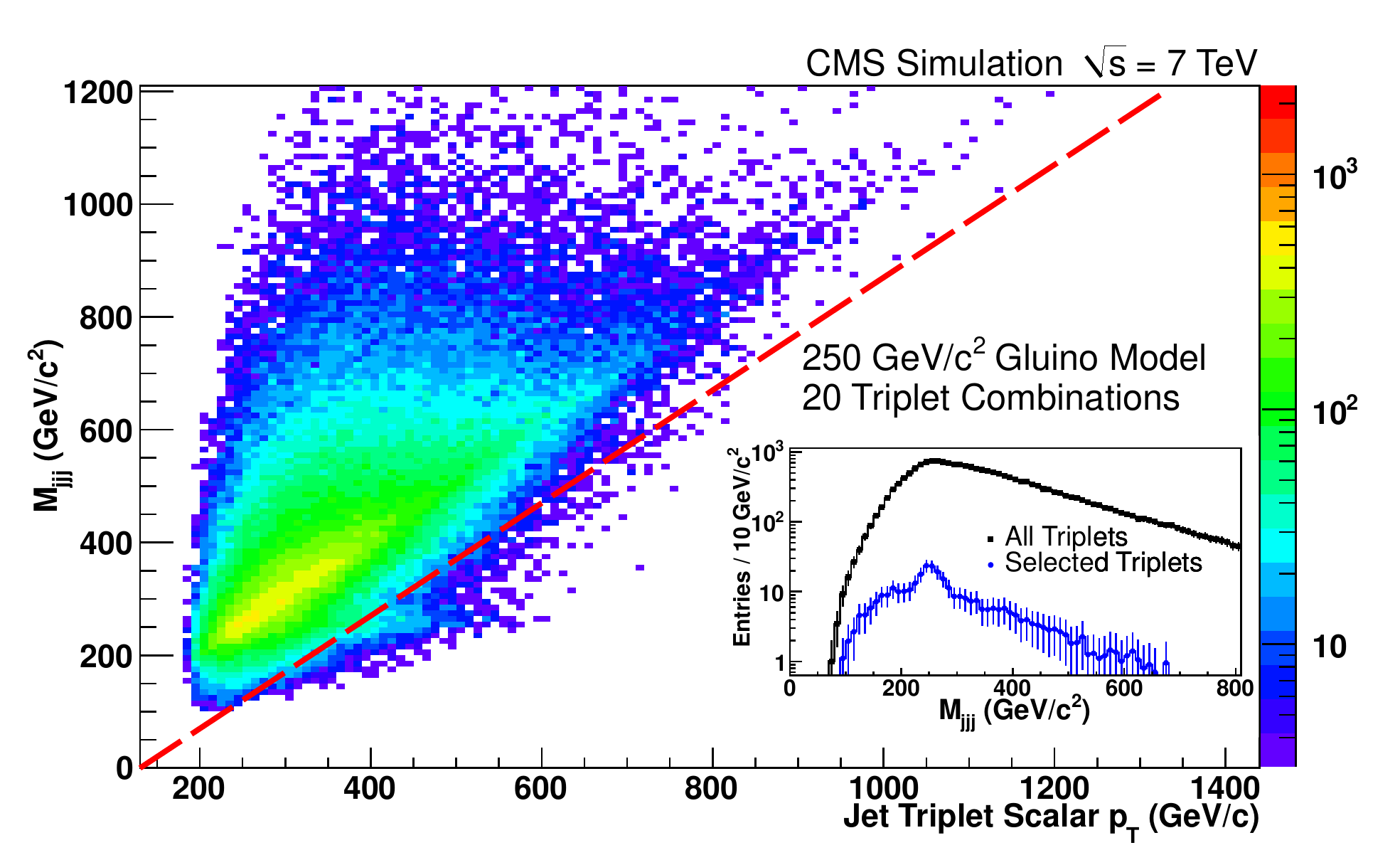}%
  \caption{\label{fig:jetresonance}Simulated triplet jet invariant mass versus the triplet scalar \pt~of all possible combinations for a 250~GeV gluino mass. All triplets falling to the right of the red dashed line pass the final selection. In the inset, the combinations before and after the selection are shown.}
\end{figure}

Both CDF (with 3.2\invfb of data)~\cite{CDF_3jet} and CMS (with 35\invpb of data)~\cite{CMS_3jet} collaborations performed a search for gluino pair production decaying into three jets. The search for two 3-jet resonances in a 6 jet final state is performed by exploiting the kinematic relationship between the jet triplet scalar \pt~and the invariant mass of the three jets. In this way, the experiments manage to reduce the combinatorics and reject the QCD multijet backgrounds, as shown in Figure~\ref{fig:jetresonance}. The complementarity between the experiments allows to fully cover a mass range from 77 to 500~GeV. With this technique, CDF excludes RPV gluino masses below 144~GeV (a $2~\sigma$ excess is found around the top mass) and CMS excludes gluino masses between 200 and 280~GeV (a $1.9~\sigma$ excess is found at 380~GeV).

%%%%%%%%%%%%%%%%%%%%%%
\section{\label{sec:summary}SUMMARY AND OUTLOOK}
Searches for supersymmetry have been carried out at the Tevatron and the LHC colliders. Thanks to the complementarity between machines, many different final states and mass ranges have been carefully scrutinized. Since no significant deviations from the SM predictions have been found, the vast parameter space available for SUSY has been substantially reduced and the most probable scenarios predicted by electroweak precision tests are now excluded or under some constraints after the new stringent limits. The question whether SUSY really exists or whether it is within the reach of the current collider experiments is becoming more relevant.

One of the great virtues of SUSY is the stabilization of the electroweak sector. The radiative corrections to the Higgs mass need of a relatively low stop mass in order to avoid too much fine tuning. This also means that gluino masses should be relatively light, since they contribute to the stop mass corrections. Thus, in order to preserve naturalness arguments for SUSY, two main scenarios can be envisioned: one is the existence of heavy squarks, intermediate gluino and light stop and gauginos, and the other is the presence of a SUSY spectrum compressed into a narrow range of masses, which would evade the current searches at colliders and would also mean that the SUSY breaking scale resides at relatively low energies. Both scenarios are still possible and will probably determine the roadmap of the searches in the coming years, at least until the LHC is able to reach the nominal 14~TeV center-of-mass energy and provide a more conclusive answer to the current open questions in our understanding of the universe.

\bigskip % extra skip inserted
\begin{acknowledgments}
The author would like to thank the organizers for their hospitality and their commitment to make this conference a successful event.

\end{acknowledgments}

\bigskip % extra skip inserted
%% Create the reference section using BibTeX:

\begin{thebibliography}{99} % Use for 10-99 references
%
%%\bibitem{accelconf-ref}
%%http://www.cern.ch/accelconf
%
%\bibitem{exampl-ref}
%\bibitem{tom} T.~Junk, Nucl. Instrum. Meth. A {\bf 434}, 435 (1999)
\bibitem{susy} H.P.~Nilles, Phys. Rep. {\bf 110}, 1 (1984), and references therein.
\bibitem{hierarchy} S. Weinberg, Phys. Rev. D {\bf 13}, 974 (1976), Phys. Rev. D {\bf 19}, 1277 (1979); E. Gildener, Phys.
Rev. D {\bf 14}, 1667 (1976); L. Susskind, Phys. Rev. D {\bf 20}, 2619 (1979); G.'t Hooft, in Recent
developments in gauge theories, Proceedings of the NATO Advanced Summer Institute, Cargese
1979, (Plenum, 1980)
\bibitem{GUT} U. Amaldi, W. de Boer, and H. Furstenau, Phys. Lett. B {\bf 260}, 447 (1991)
\bibitem{mssm} S.P.~Martin, {\tt arXiv:hep-ph/9709356v6}
\bibitem{CMSSM} G. L. Kane et al., Phys. Rev. D {\bf 49}, 6173-6210 (1994), {\tt arXiv:hep-ph/9312272}
\bibitem{GMSB} M. Dine et al., Nucl.Phys. B {\bf 189}, 575 (1981)
\bibitem{LEPsusy} ALEPH+DELPHI+L3+OPAL Collabs. SUSY working group page, \url{http://lepsusy.web.cern.ch/lepsusy}, and references therein.
\bibitem{Rparity} R. Barbier et al., Phys. Rept. {\bf 420}, 1 (2005) {\tt arXiv:hep-ph/0406039}
\bibitem{wwwexp} CDF:~\url{http://www-cdf.fnal.gov/physics/exotic/exotic.html};\\ D\O: \url{http://www-d0.fnal.gov/Run2Physics/WWW/results/np.htm};\\ ATLAS:~\url{https://twiki.cern.ch/twiki/bin/view/AtlasPublic/SupersymmetryPublicResults};\\ CMS:~\url{https://twiki.cern.ch/twiki/bin/view/CMSPublic/PhysicsResultsSUS}
\bibitem{ATLAS_0lep} ATLAS Collab., {\tt arXiv:1109.6572 [hep-ex]}
\bibitem{ATLAS_multijets} ATLAS Collab., {\tt arXiv:1110.2299 [hep-ex]}
\bibitem{CMS_alphaT} CMS Collab., {\tt arXiv:1109.2352 [hep-ex]}
\bibitem{CMS_Razor} CMS Collab., {\tt arXiv:1107.1279 [hep-ex]}
\bibitem{alphaT} L. Randall and D. Tucker-Smith, Phys. Rev. Lett. {\bf 101}, 221803 (2008)
\bibitem{Razor} C. Rogan, {\tt arXiv:1006.2727 [hep-ph]}
\bibitem{ATLAS_1lep} ATLAS Collab., {\tt arXiv:1109.6606 [hep-ex]}
\bibitem{ATLAS_MM} ATLAS Collab., Eur. Phys. J. C {\bf 71} 1577 (2011), {\tt arXiv:1012.1792 [hep-ex]}
\bibitem{CMS_1lep} CMS Collab., PAS-SUS-11-015
\bibitem{CDF_SS} CDF Collab., CDF/PHYS/EXO/PUBLIC/10464
\bibitem{CMS_SS} CMS Collab., PAS-SUS-11-010
\bibitem{CMS_OS} CMS Collab., PAS-SUS-11-011
\bibitem{ATLAS_2l} ATLAS Collab., {\tt arXiv:1110.6189 [hep-ex]}
\bibitem{D0_trilep} D\O~ Collab., Phys. Lett. B {\bf 680}, 34 (2009) {\tt arXiv:0901.0646 [hep-ex]}
\bibitem{CDF_trilep} CDF Collab., CDF/PUB/EXOTIC/PUBLIC/10636
\bibitem{CDF_sbottom} CDF Collab., Phys. Rev. Lett. {\bf 105}, 081802 (2010)
\bibitem{D0_sbottom} D\O~ Collab., Phys. Lett. B {\bf 693}, 95 (2010), {\tt arXiv:1005.2222 [hep-ex]}
\bibitem{D0_stop} D\O~ Collab., Phys. Lett. B {\bf 696}, 321 (2011), {\tt arXiv:1009.5950 [hep-ex]}
\bibitem{ATLAS_glsb} ATLAS Collab., ATLAS-CONF-2011-098
\bibitem{ATLAS_glst} ATLAS Collab., ATLAS-CONF-2011-130
\bibitem{SPS8} B.C. Allanach {\it et al.}, Eur. Phys. J. C {\bf 25} 113 (2002), {\tt arXiv:hep-ph/0202233}
\bibitem{D0_photons} D\O~ Collab., Phys. Rev. Lett. {\bf 105}, 221802 (2010), {\tt arXiv:1008.2133 [hep-ex]}
\bibitem{CDF_photons} CDF Collab., Phys. Rev. Lett. {\bf 104}, 011801 (2010)
\bibitem{GGM} J.T. Ruderman and D. Shih, {\tt arXiv:1103.6083 [hep-ph]}
\bibitem{ATLAS_photons} ATLAS Collab., {\tt arXiv:1111.4116 [hep-ex]}
\bibitem{CMS_photons} CMS Collab., PAS-SUS-11-009
\bibitem{D0_staunu} D\O~ Collab.,  Phys. Rev. Lett. {\bf 105}, 191802 (2010), {\tt arXiv:1007.4835 [hep-ex]}
\bibitem{ATLAS_staunu} ATLAS Collab., {\tt arXiv:1109.3089 [hep-ex]}
\bibitem{CDF_3jet} CDF Collab., Phys. Rev. Lett. {\bf 107}, 042001 (2011)
\bibitem{CMS_3jet} CMS Collab., Phys. Rev. Lett. {\bf 107}, 101801 (2011), {\tt arXiv:1107.3084 [hep-ex]}

%%\bibitem{templates-ref}
%%http://www.cern.ch/accelconf/templates.html
%
\end{thebibliography}
%\begin{thebibliography}{9}   % Use for  1-9  references

\end{document}